\documentclass[]{jetp} 
\usepackage{xcolor}

\begin{document}
\English

\title{Flow-Regulated Suprathermal Particle Acceleration in Weakly Collisional Astrophysical Plasmas}
\rtitle{Flow-Regulated Particle Acceleration\dots}

\author{Ji-Hoon}{Ha} 
\affiliation{Korea Astronomy and Space Science Institute, 34055, Daejeon, South Korea}
\email{hjhspace223@gmail.com}

\author{Elena S.}{Volnova} 
\affiliation{Institute for Basic Science, 34126, Daejeon, South Korea}

\abstract{We investigate the formation of suprathermal particle populations in weakly collisional plasmas using a one-dimensional Fokker-Planck framework. 
A key element of this work is the introduction of a systematic velocity-space drift term that represents net energization relative to a background streaming flow. 
This term provides a minimal phenomenological description of competing relaxation and acceleration processes, enabling the incorporation of large-scale plasma dynamics into kinetic particle evolution.
The model further includes spatial advection, velocity-space diffusion associated with wave-particle interactions, and collisional relaxation. 
To explore the role of time-dependent large-scale plasma dynamics, we consider three representative temporal profiles of the streaming velocity: accelerating, decelerating, and steady flows. 
We find that velocity-space diffusion primarily governs the formation of suprathermal tails, while the streaming-induced drift regulates their efficiency and spectral properties. 
In particular, the overall fraction of suprathermal particles depends only weakly on the detailed temporal evolution of the flow and is largely controlled by the time-averaged streaming velocity. 
These results suggest that large-scale streaming motions can be incorporated as an effective systematic energization mechanism in weakly collisional plasmas, providing a minimal and flexible framework applicable to a broad range of space and astrophysical environments.}

\maketitle


\section{Introduction}\label{sec1}

Suprathermal particle populations, characterized by velocity distributions extending beyond a Maxwellian core, are commonly observed in a variety of weakly collisional space plasmas, including the solar wind, planetary magnetospheres, and the heliosheath \cite{montgomery1968, feldman1975, pilipp1987, maksimovic1997a, maksimovic1997b, maksimovic2005, marsch1982, marsch2006, pierrard2010}. It has been shown that the electron distribution in the solar wind consists of three distinct components: a thermal core, a suprathermal halo, and a suprathermal strahl \cite{pilipp1987, maksimovic1997a, maksimovic2005, marsch2006,stverak2009, pierrard2010}. Observational studies primarily examine the characteristics of suprathermal electrons in the halo, including the spectral slope of the distribution function, number density, temperature, plasma beta, and temperature anisotropy \cite{lazar2020}. The scattering of suprathermal electrons by plasma waves is also investigated, with physical arguments focusing on kinetic turbulence involving whistler waves \cite{pagel2007}. Observational evidence for the presence of suprathermal protons has also been presented \cite{marsch1982, marsch2006}. These nonthermal populations play crucial roles in energy transport, wave generation, and particle acceleration processes \cite{marsch2006, pierrard2010}, yet their formation mechanisms remain incompletely understood.

One of the prevailing theories suggests that suprathermal particles originate from a combination of resonant wave-particle interactions and collisional processes acting on large-scale plasma flows. For example, the solar wind provides a natural laboratory where turbulence persists over extended distances and timescales, continually shaping the velocity distributions of ions and electrons \cite{marsch2006}. Kinetic simulations, including those employing the Particle-in-Cell (PIC) method, have been extensively performed to examine the properties of waves responsible for wave-particle interactions \cite{ryu2007, ohira2008, gary2011, umeda2012, shaaban2018, lazar2019, lazar2022, kim2020}. Relevant plasma instabilities are driven by free energy sources such as beam-plasma interactions and temperature anisotropy. In particular, beam-driven instabilities in various environments, including shocks and turbulence, significantly enhance particle acceleration \cite{ryu2007, ohira2008, umeda2012, shaaban2018, kim2020}. In addition, inhomogeneous plasma structures can generate drift motions that act as free energy sources for plasma instabilities, such as the lower-hybrid drift instability (LHDI), which can also contribute to particle acceleration through wave-particle interactions \cite{lavorenti2021, ha_volnova2025}. Furthermore, it has been shown that the time evolution of the suprathermal electron distribution function is associated with wave-particle interactions due to whistler waves driven by temperature anisotropy \cite{gary2011, lazar2019, lazar2022}. While kinetic simulations contribute to a better understanding of plasma waves and associated wave-particle interactions, they are insufficient to fully capture the evolution over the dynamic timescales characteristic of space and astrophysical environments.

To describe the acceleration process mediated by wave-particle interactions during the dynamical evolution of the system, the Fokker-Planck equation is widely used due to its ability to incorporate both diffusive and collisional effects within a kinetic framework \cite{schlickeiser2002}. Specifically, wave-particle interactions mediated by Langmuir and whistler turbulence have been extensively studied \cite{pierrard2011, zank2014, kim2015, tang2018, tang2022}, while Coulomb collisional effects are simultaneously considered \cite{tang2024, ha2025}. For example, in the space environments, it has been shown that the wave-particle interaction term dominates over the Coulomb collision terms at large radial distances from the Sun, and the steady-state solution of the Fokker-Planck equation deviates from a Maxwellian distribution due to the dominant wave-particle interaction driven by whistler turbulence \cite{shizgal2007, pierrard2011}. Regarding proton acceleration, the role of kinetic Alfv\'en waves has been rigorously considered \cite{rudakov2012, pierrard2013, voitenko2013, choi2021, ha2024}. The acceleration efficiency is significantly influenced by the presence of suprathermal protons in the initial distribution function and the plasma beta of the system \cite{ha2024}.

While previous studies have primarily focused on steady-state solutions or prescribed background turbulence, fewer have examined how large-scale, time-dependent plasma flows influence the formation of suprathermal particle populations within a kinetic framework. 
In conventional Fokker-Planck treatments, external effects are often represented through simplified, time-independent forces such as static electric fields or gravitational acceleration \cite{pierrard2011, pierrard2013, tang2024, ha2025}. 
However, time-dependent bulk flows are ubiquitous in space and astrophysical plasmas, including the solar wind, galaxy clusters, relativistic jets, and accretion flows, where plasma motions undergo acceleration, compression, and deceleration over a wide range of scales \cite{zirker1977, komissarov2007, hong2014, yuan2014, ha2018}. 
More generally, the Fokker--Planck equation admits a systematic velocity-space drift term that describes net energization or relaxation processes acting on particles. 
In weakly collisional plasmas, such a drift can arise from competing effects, including scattering-driven relaxation toward the local flow frame and energization by large-scale electric fields or wave--particle interactions. 
In this context, large-scale streaming motions provide a natural means to parameterize net systematic energization relative to the background flow, consistent with observational evidence that particle acceleration is enhanced during periods of elevated solar wind speed and correlated with bulk flow properties \cite{baker1990}.

In this work, we model the influence of streaming flows through an effective velocity-space drift term in the time-dependent Fokker-Planck equation. 
This approach enables the incorporation of both the magnitude and temporal evolution of the background flow in a physically transparent manner, without requiring detailed knowledge of the underlying microphysical acceleration processes. 
To investigate the impact of evolving flows, we solve a one-dimensional Fokker-Planck equation that includes spatial advection, velocity-space diffusion due to wave-particle interactions, and collisional relaxation. 
We consider three representative models of the streaming velocity-accelerating, decelerating, and steady profiles in order to examine how different temporal behaviors of large-scale flows influence the formation and evolution of suprathermal particle populations.

The remainder of this paper is organized as follows. Section~\ref{sec2} describes the governing equations and numerical implementation, Section~\ref{sec3} presents the results for different streaming flow profiles, and Section~\ref{sec4} provides a brief summary.

\section{Model description}\label{sec2}

\subsection{Phenomenological description of velocity-space drift in streaming plasmas}\label{subsec2.1}

In addition to diffusive acceleration, we introduce a systematic velocity-space drift term that accounts for net energization relative to the background streaming flow. 
Rather than representing a literal microscopic force acting on individual particles, this term should be interpreted as an effective, coarse-grained description of competing physical processes in a weakly collisional plasma.

In general, particles interacting with a background flow tend to relax toward the local flow frame due to scattering processes, which can be described by a relaxation term of the form
\begin{equation}
A_{\rm rel}(v) = -\frac{v - v_{\rm st}}{\tau_{\rm rel}},
\label{eq1}
\end{equation}
where $\tau_{\rm rel}$ is the characteristic relaxation timescale. 
At the same time, large-scale electric fields and wave-particle interactions can give rise to systematic energization in velocity space. 
In the present framework, this effect is not modeled explicitly, but is instead represented through an effective drift term of the form
\begin{equation}
A_{\rm acc}(v) = +\frac{v - v_{\rm st}}{\tau_{\rm acc}},
\label{eq2}
\end{equation}
where $\tau_{\rm acc}$ characterizes the timescale of net energization.
A natural justification for the linear dependence on $(v - v_{\rm st})$ is that it represents the leading-order term in a velocity-space expansion around the local streaming frame. 
Near $v \approx v_{\rm st}$, higher-order contributions are expected to be subdominant, and the dominant systematic effect can be captured by retaining only the linear term in $(v - v_{\rm st})$. 
This approximation provides a minimal closure for describing net energization relative to the background flow.

The net effect of these competing processes can be expressed as a combined drift coefficient
\begin{equation}
A_{\rm eff}(v) = \left( \frac{1}{\tau_{\rm acc}} - \frac{1}{\tau_{\rm rel}} \right) (v - v_{\rm st}).
\label{eq3}
\end{equation}
In the regime where energization dominates over relaxation ($\tau_{\rm acc} < \tau_{\rm rel}$), this results in an effective anti-drag term,
\begin{equation}
A_{\rm eff}(v) \approx +\frac{v - v_{\rm st}}{\tau_0},
\label{eq4}
\end{equation}
which corresponds to the form adopted in Equation \eqref{eq2}.
Here, $\tau_0$ is an effective timescale that characterizes the net rate of systematic energization in velocity space. 
From Equation~\eqref{eq3}, it can be expressed as $\tau_0^{-1} = \tau_{\rm acc}^{-1} - \tau_{\rm rel}^{-1}$, indicating that it represents the competition between acceleration and relaxation processes. 
The convective acceleration term used in this work should be understood as a minimal phenomenological closure that captures net systematic energization in the presence of a streaming background, rather than a direct representation of a specific microscopic force. 
Accordingly, it does not imply direct particle energization at the single-particle level, but instead represents the evolution of the distribution function in velocity space arising from collective processes.

\begin{figure}[t]
    \centering
    \includegraphics[width=1\linewidth]{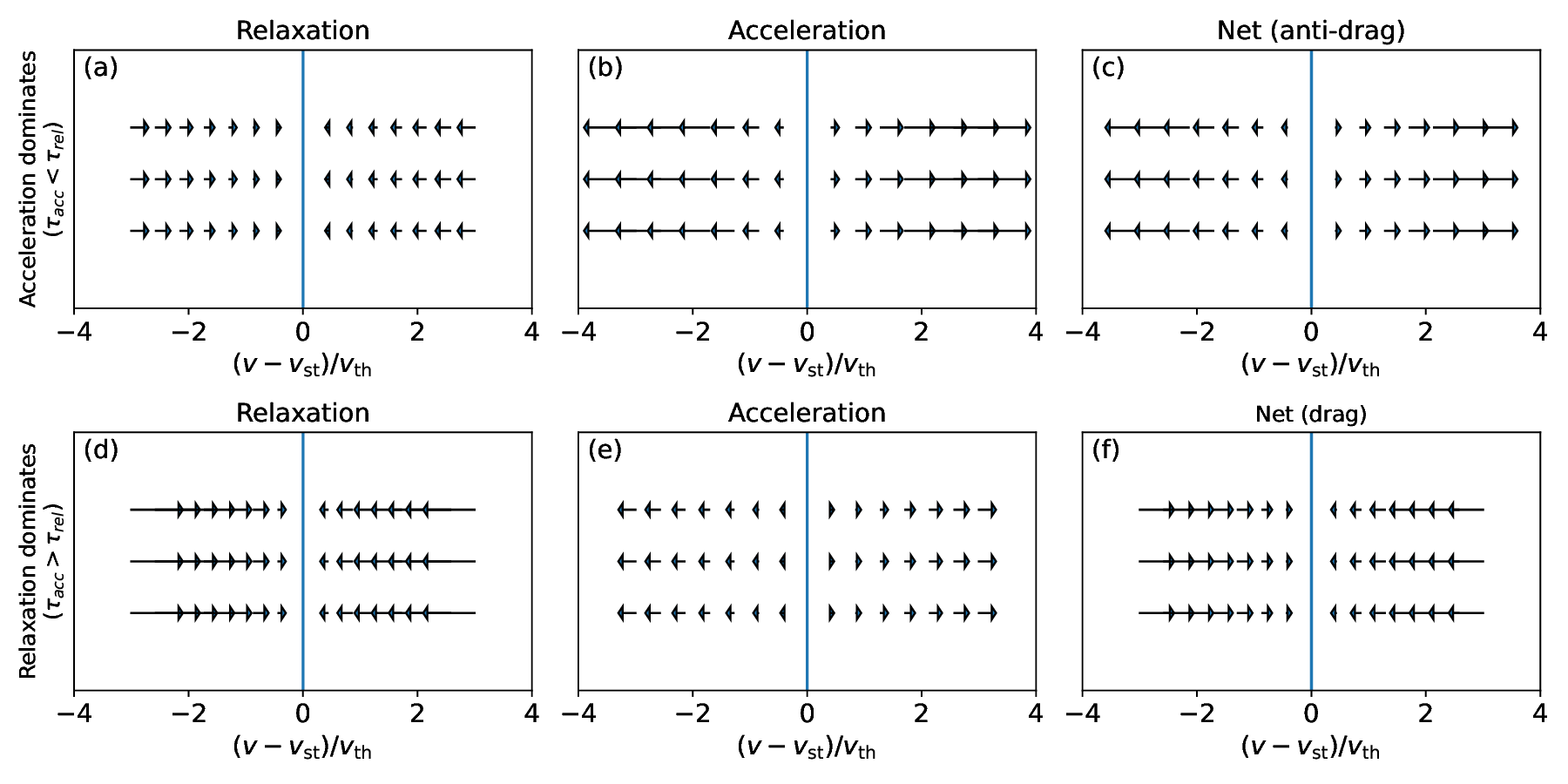}
    \caption{Schematic illustration of competing velocity-space processes in a streaming plasma. Panels (a)-(c) show the case where energization dominates over relaxation ($\tau_{\rm acc} < \tau_{\rm rel}$), leading to an effective anti-drag. Panels (d)-(f) show the opposite regime, where relaxation dominates and particles drift toward the streaming velocity $v_{\rm st}$. This schematic provides a physical interpretation of the effective drift term introduced in Section~\ref{subsec2.1}.}
    \label{fig:f1}
\end{figure}

The physical interpretation of the effective drift term is illustrated schematically in Figure~\ref{fig:f1}. 
In a weakly collisional plasma, particles experience competing processes: scattering tends to relax particles toward the local streaming frame, while large-scale electric fields and wave--particle interactions can drive systematic energization away from it. 
When relaxation dominates ($\tau_{\rm acc} > \tau_{\rm rel}$), particles drift toward $v_{\rm st}$, resulting in an effective drag in velocity space. 
In contrast, when energization dominates ($\tau_{\rm acc} < \tau_{\rm rel}$), the net effect becomes an anti-drag, leading to outward transport in velocity space and the formation of suprathermal tails. 
This competition naturally gives rise to the effective drift term adopted in this work.

\subsection{One-dimensional Fokker-Planck equation}\label{subsec2.2}

We employ a Maxwellian distribution function as the initial velocity distribution:
\begin{equation}
f_0(x,v,t=0) = \frac{N_{0}(x)}{\sqrt{2\pi v_{\text{th}}^2}} \exp\left(-\frac{v^2}{2v_{\text{th}}^2}\right),
\label{eq5}
\end{equation}
where $v_{\text{th}}$ is the initial thermal velocity, and $N_{0}(x)$ is a normalization factor depending on the spatial coordinate $x$. For a homogeneous plasma system, the normalization factor simplifies to a constant, $N_{0}(x) = N_0$.

In a one-dimensional domain with periodic boundary conditions, we solve the following Fokker-Planck equation:
\begin{align}
\frac{\partial f(x,v,t)}{\partial t} 
&+ v \frac{\partial f(x,v,t)}{\partial x} 
+ \frac{\partial}{\partial v} \left[ A_{\rm eff}(v) f(x,v,t) \right] \nonumber \\
&+ \frac{\partial}{\partial v} \left( \frac{D(v)}{\tau_0^2} \frac{\partial f(x,v,t)}{\partial v} \right) 
+ \frac{\partial}{\partial v} \left( \frac{v}{\tau_{\text{coll}}} \left[ f(x,v,t) - f_{\text{eq}}(x,v,\tau_{\text{eq}}) \right] \right) = 0.
\label{eq6}
\end{align}
where $\tau_0$ is an effective timescale that characterizes the net rate of systematic energization in velocity space, as defined in Section~\ref{subsec2.1}. 
It represents the balance between acceleration and relaxation processes and serves as a reference timescale for the evolution of the system. $v_{\text{st}}(t)$ denotes the time-dependent bulk streaming velocity of the background plasma, $D(v)$ is the diffusion coefficient, $\tau_{\text{coll}}$ is the characteristic collisional timescale, and $f_{\text{eq}}$ is a Maxwellian equilibrium distribution representing the relaxed state due to dominant collisional effects. Further details of the equation are provided in the following paragraphs.

We note that the velocity-space drift term in Equation~\eqref{eq6} is written in conservative form, 
$\partial/\partial v \left[ A_{\rm eff}(v) f \right]$, so that the equation preserves particle number. 
Indeed, integrating Equation~\eqref{eq6} over velocity space, the velocity-space drift, diffusion, and collisional relaxation terms reduce to boundary contributions. 
Assuming that the distribution function decreases sufficiently rapidly at large velocities, these boundary terms vanish, and the total particle number density is conserved in the absence of explicit source or sink terms. 
This conservative form is fully consistent with the standard Fokker--Planck description of systematic transport in velocity space.

The time evolution of the distribution function is governed by advection in both physical and velocity space, as well as velocity diffusion. 
In particular, the velocity-space advection term, represented by the third term in Equation~\eqref{eq6}, includes the time-dependent bulk streaming velocity of the background plasma, $v_{\text{st}}(t)$ (see Section~\ref{subsec2.3} for more details). 
The factor $A_{\rm eff}(v) \approx (v - v_{\text{st}}(t))/\tau_0$ defines a systematic drift in velocity space that describes net energization relative to the background flow. 
Its physical interpretation as an effective, coarse-grained description of competing processes is discussed in Section~\ref{subsec2.1}. 

Along with systematic energization represented by the drift term, stochastic acceleration in velocity space is modeled through a diffusion coefficient, $D(v)$, which is taken in the form
\begin{equation}
D(v) \approx D_0 \left(\frac{v}{v_{\text{th}}}\right)^{\frac{s-1}{3}}\exp\left(-\frac{v}{v_{\text{max}}}\right),
\label{eq7}
\end{equation}
where $v_{\text{max}}$ denotes a characteristic upper velocity beyond which the efficiency of resonant wave-particle interactions is suppressed. In this work, we adopt $v_{\text{max}} = 10 v_{\text{th}}$ as a representative cutoff scale; the results are not sensitive to the exact choice of this parameter.
The sublinear velocity dependence is motivated by kinetic models of electron transport in the presence of whistler turbulence, in which the velocity dependence of the diffusion coefficient is determined by the spectral properties of the turbulent wave field and resonance conditions \cite{steinacker1992, pierrard2011, tang2018, tang2022, tang2024, ha2025}. 
In such models, the effective scaling of the diffusion coefficient can be expressed as a weak power-law in velocity, corresponding to spectral indices in the range $s \sim 3/2$-2. 
In this work, we adopt $s = 2$ as a representative baseline value for the underlying turbulence spectrum and use $s = 3/2$ for comparison. We have verified that moderate variations of $s$ within this range do not qualitatively affect the main results of this paper, indicating that our conclusions are robust with respect to the specific choice of the diffusion scaling. 
The exponential factor is introduced phenomenologically to represent the finite resonance range, ensuring that stochastic acceleration becomes inefficient above a characteristic velocity scale. 
Accordingly, the adopted form should be regarded as a physically motivated minimal model: the $v^{(s-1)/3}$ dependence reflects the literature-based scaling of wave-particle diffusion, while the exponential cutoff captures the expected suppression of resonant scattering at sufficiently high velocities.

To incorporate the effects of particle collisions in the kinetic equation, we adopt the Bhatnagar-Gross-Krook (BGK) approximation, which replaces the full Boltzmann collision operator with a simplified relaxation term \cite{pan2018}. 
In this approach, the distribution function $f$ relaxes toward a local Maxwellian equilibrium distribution $f_{\text{eq}}$ over a characteristic timescale $\tau_{\text{coll}}$, such that the collisional term is written as $-(f - f_{\text{eq}})/\tau_{\text{coll}}$.
We emphasize that this treatment is intended as a minimal phenomenological description of collisional relaxation in a weakly collisional plasma. 
While a more rigorous description would require a Fokker-Planck collision operator for Coulomb interactions, the BGK model captures the essential effect of collisions-namely, the tendency of the particle distribution to relax toward local thermodynamic equilibrium-without introducing additional complexity associated with velocity-dependent collision terms. 
Given that the primary focus of this work is on the interplay between stochastic acceleration and systematic energization, rather than the detailed microphysics of Coulomb collisions, the BGK approximation provides a computationally efficient and physically adequate description of collisional effects.

To illustrate the collisional regime relevant for space and astrophysical plasmas, we compare a characteristic dynamical timescale with the electron--electron Coulomb collision timescale using representative parameters of the solar wind as a concrete example \cite{bruno2013}. 
The dynamical timescale can be estimated as
\begin{equation}
\tau_{\rm dyn} \sim \frac{L_{\rm dyn}}{V_{\rm SW}} \sim \frac{10^6~{\rm km}}{500~{\rm km/s}} \sim 2 \times 10^3~{\rm s},
\label{eq8}
\end{equation}
where $V_{\rm SW} \sim 500~{\rm km/s}$ is the bulk flow speed and $L_{\rm dyn} \sim 10^6~{\rm km}$ represents a characteristic large-scale dynamical length. 
For typical solar wind conditions ($n_e \sim 5~{\rm cm^{-3}}$, $T_e \sim 10^5~{\rm K}$, ${\rm ln}\Lambda \sim 10$), the electron--electron Coulomb collision timescale is given by
\begin{equation}
\tau_{\rm ee} = \frac{3\sqrt{m_e}(k_B T_e)^{3/2}}{4\sqrt{2}\pi n_e e^4 {\rm ln} \Lambda} \sim 3 \times 10^4~{\rm s}.
\label{eq9}
\end{equation}
The resulting ratio $\tau_{\rm ee}/\tau_{\rm dyn} \sim 15$ indicates that the plasma is weakly collisional, allowing the formation and persistence of suprathermal particle populations. 
Such a separation of timescales is commonly expected in a wide range of space and astrophysical plasmas, justifying the use of a weakly collisional kinetic description in this work. 
Motivated by this hierarchy, we adopt $\tau_{\text{coll}} \approx 10\,\tau_0$ as a representative parameter choice to capture the dynamical evolution in weakly collisional environments. We additionally examined a more weakly collisional case with $\tau_{\rm coll} \approx 100\,\tau_0$ to test the robustness of the results in a more weakly collisional regime.

\subsection{Modeling of time-dependent streaming flow}
\label{subsec2.3}
The justification for adopting a time-varying streaming flow model is grounded in the inherently dynamic nature of space and astrophysical plasmas, where large-scale flows often exhibit significant temporal and spatial variations. 
Such variability can arise from evolving background conditions, interactions between distinct plasma streams, or large-scale dynamical processes that modify the bulk flow velocity over time.
A representative example is the solar wind, where streams originating from coronal holes undergo acceleration and interact with the ambient plasma, producing compression and stream interaction regions characterized by time-dependent variations in flow velocity \cite{zirker1977, garton2018, jian2006, richardson2018}.
In galaxy clusters, time-dependent bulk flows naturally arise in multiple physical contexts. 
For example, accretion of filamentary gas onto clusters generates large-scale inflows that accelerate and form accretion shocks \cite{hong2014}, while merger events drive complex shock structures and turbulent flows that involve both compression and subsequent deceleration in downstream regions \cite{ha2018}.
In relativistic or mildly relativistic outflows such as AGN jets, the bulk flow velocity can vary significantly due to acceleration near the launching region, as well as deceleration caused by interaction with the ambient medium or internal dissipation processes \cite{komissarov2007, marscher2014}. 
Similarly, in accretion flows onto compact objects, plasma is subject to gravitational acceleration and turbulent interactions, leading to time-dependent variations in the flow velocity \cite{balbus1991,narayan1994,yuan2014,ha2026}. 
These examples suggest that time-varying streaming flows are a generic feature of weakly collisional plasmas, supporting the use of simplified time-dependent models such as those adopted in this work.

\begin{figure}[t]
    \centering
    \includegraphics[width=0.7\linewidth]{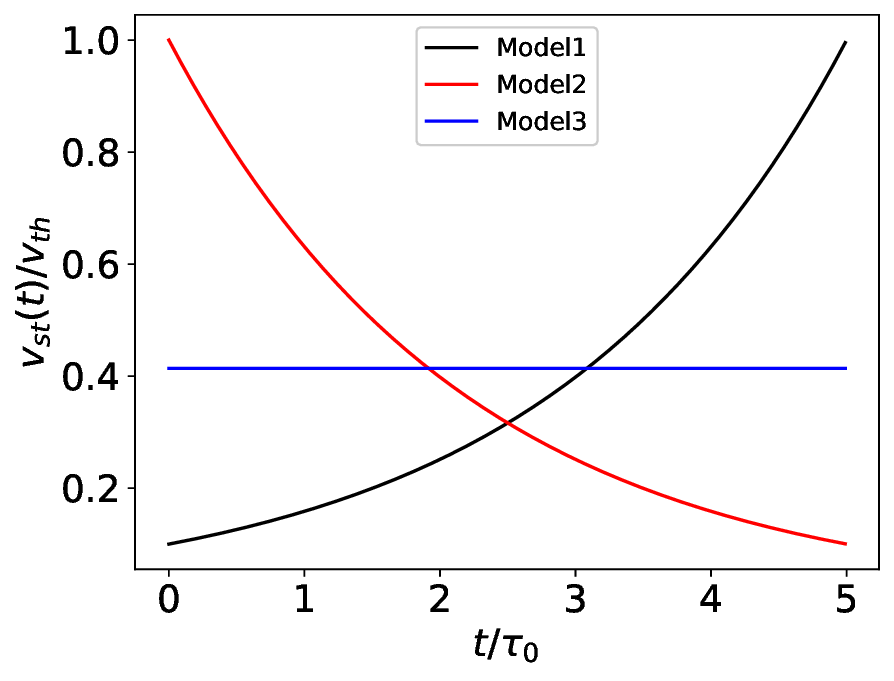}
    \caption{Three different models of time-varying streaming flow. Model 1 and Model 2 assume accelerating and decelerating streaming flows, respectively, while Model 3 assumes a constant streaming velocity.}
    \label{fig:f2}
\end{figure}

Motivated by these considerations, we model the streaming velocity as a time-dependent function with a characteristic timescale $\tau_0$, which provides a simple yet physically motivated representation of evolving large-scale flows. 
While the specific example above is drawn from the solar wind, the adopted formulation is intended to capture generic features of time-dependent bulk motion in weakly collisional plasmas.
To capture different modes of flow evolution, we consider three representative models of time-varying streaming velocity. 
The first model describes an accelerating flow,
\begin{equation}
v_{\text{st,1}}(t) = v_{\text{i,1}}~\exp\left(\Gamma_{\text{st}}t\right),
\label{eq10}
\end{equation}
while the second model represents a decelerating flow,
\begin{equation}
v_{\text{st,2}}(t) = v_{\text{i,2}}~\exp\left(-\Gamma_{\text{st}}t\right).
\label{eq11}
\end{equation}
Here, $\Gamma_{\text{st}}$ characterizes the rate of change of the bulk flow velocity, and we adopt $\Gamma_{\text{st}} = \ln(10)/T$ with $T = 5\tau_0$ as a representative choice.
For comparison, we also consider a constant-flow model defined by the time-averaged streaming velocity,
\begin{equation}
v_{\text{st,3}}(t) = \left<v_{\text{st,1}}(t)\right> = \left<v_{\text{st,2}}(t)\right>,
\label{eq12}
\end{equation}
where
\begin{equation}
\left<v_{\text{st,1}}(t)\right> = \frac{1}{T} \int_{0}^{T} v_{\text{st,1}}(t) \, dt = \frac{v_{\text{i,1}}}{T}\frac{e^{\Gamma_{\text{st}}T}-1}{\Gamma_{\text{st}}},
\label{eq13}
\end{equation}
\begin{equation}
\left<v_{\text{st,2}}(t)\right> = \frac{1}{T} \int_{0}^{T} v_{\text{st,2}}(t) \, dt = \frac{v_{\text{i,2}}}{T}\frac{1-e^{-\Gamma_{\text{st}}T}}{\Gamma_{\text{st}}}.
\label{eq14}
\end{equation}
These three models represent accelerating, decelerating, and steady large-scale flows, respectively, and serve as idealized descriptions of time-dependent bulk motion in plasmas. 
The temporal evolution of the streaming velocity for each model is illustrated in Figure~\ref{fig:f2}. 
Hereafter, we refer to the accelerating and decelerating cases as Model 1 and Model 2, respectively, while Model 3 denotes the constant-flow case.
The characteristic amplitude of the streaming velocity is chosen to be a fraction of the thermal velocity, comparable to typical turbulent velocities expected in astrophysical plasmas such as the intracluster medium \cite{brunetti2014}.

\subsection{Numerical scheme}
\label{subsec2.4}

For solving the equation numerically, we adopt the Crank-Nicolson method for time integration \cite{neena2022}, which provides a balance between numerical stability and accuracy. Spatial and velocity derivatives are discretized using second-order accurate central difference schemes. Specifically, the first-order derivatives in space and velocity are approximated as follows:
\begin{equation}
\left. \frac{\partial f}{\partial x} \right|_i \approx \frac{f_{i+1} - f_{i-1}}{2 \Delta x},
\label{eq15}
\end{equation}
\begin{equation}
\left. \frac{\partial f}{\partial v} \right|_j \approx \frac{f_{j+1} - f_{j-1}}{2 \Delta v},
\label{eq16}
\end{equation}
where $i$ and $j$ denote the indices corresponding to the spatial and velocity grids, respectively.

The velocity-space drift term is written in conservative form as
\begin{equation}
\frac{\partial}{\partial v} \left[ A_{\rm eff}(v)\,f \right].
\label{eq17}
\end{equation}
For numerical implementation, this term is evaluated by first defining the velocity-space flux
\begin{equation}
J_j = A_{{\rm eff},j} f_j,
\label{eq18}
\end{equation}
and then computing its divergence using a central difference scheme,
\begin{equation}
\left. \frac{\partial}{\partial v} \left( A_{\rm eff} f \right) \right|_j 
\approx \frac{J_{j+1} - J_{j-1}}{2\Delta v}.
\label{eq19}
\end{equation}
This formulation ensures that the drift term is treated in a conservative manner, consistent with the standard Fokker--Planck description.

The velocity-space diffusion term,
\begin{equation}
\frac{\partial}{\partial v} \left[ \frac{D(v)}{\tau_0^2} \frac{\partial f}{\partial v} \right],
\label{eq20}
\end{equation}
is evaluated using a two-step procedure. First, the second derivative of $f$ with respect to velocity at the grid point $j$ is approximated as
\begin{equation}
\left. \frac{\partial^2 f}{\partial v^2} \right|_j \approx \frac{f_{j+1} - 2f_j + f_{j-1}}{\Delta v^2}.
\label{eq21}
\end{equation}
To incorporate Crank-Nicolson time averaging, we evaluate the diffusion term at both the current time level $n$ and the next level $n+1$, and then compute their average:
\begin{align}
\frac{\partial}{\partial v} \left[\frac{D(v)}{\tau_0^2} \frac{\partial f}{\partial v} \right] &\approx \frac{1}{2\tau_0^2} \left( \frac{D_{j+1} - D_{j-1}}{2 \Delta v} \right) \left( \frac{f_{j+1}^n - f_{j-1}^n}{2 \Delta v} \right) \notag \\
&+ \frac{1}{2\tau_0^2} D_j \left( \frac{f_{j+1}^n - 2f_j^n + f_{j-1}^n}{\Delta v^2} \right) \notag \\
&+ \frac{1}{2\tau_0^2} \left( \frac{D_{j+1} - D_{j-1}}{2 \Delta v} \right) \left( \frac{f_{j+1}^{n+1} - f_{j-1}^{n+1}}{2 \Delta v} \right) \notag \\
&+ \frac{1}{2\tau_0^2} D_j \left( \frac{f_{j+1}^{n+1} - 2f_j^{n+1} + f_{j-1}^{n+1}}{\Delta v^2} \right).
\label{eq22}
\end{align}

For the numerical stability test, we assume spatial, velocity, and temporal resolutions of $dx = 10^{-3}L$, $dv = 10^{-3}v_{\text{max}}$, and $dt = 10^{-3}\tau_0$, respectively. Here, $L$ denotes the spatial domain size, $v_{\text{max}}$ is the maximum velocity in the system, and $\tau_0$ is the effective timescale introduced in Section \ref{subsec2.1}.

\section{Generation of suprathermal particles}\label{sec3}

\begin{figure}[t]
    \centering
    \includegraphics[width=1\linewidth]{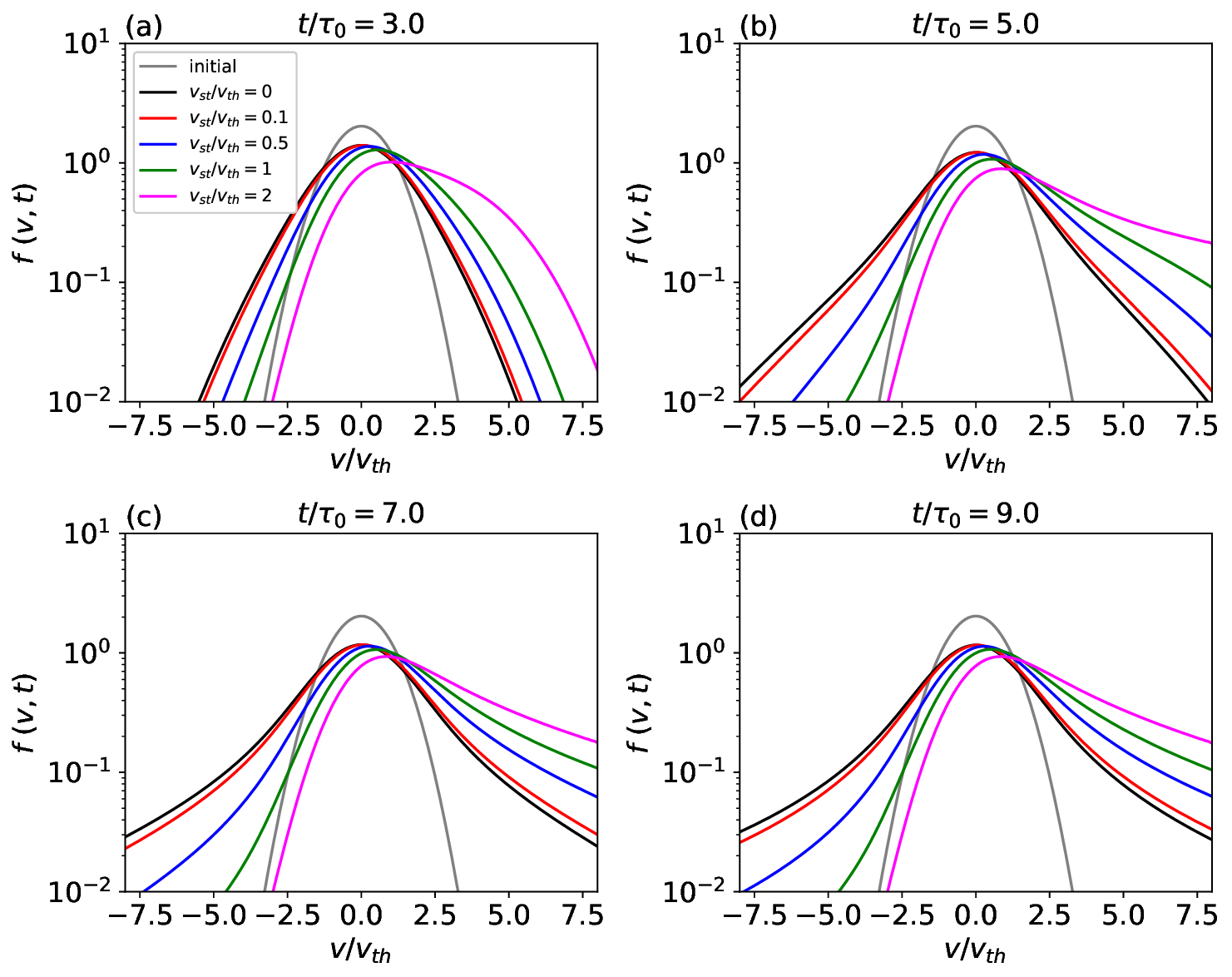}
    \caption{Time evolution of the velocity distribution function for different streaming velocities. The gray line denotes the initial distribution. A constant streaming velocity (Model 3) is assumed. The fiducial model adopts a diffusion coefficient with $s=2$ and a collisional timescale $\tau_{\rm coll} = 10\,\tau_0$.}
    \label{fig:f3}
\end{figure}

We primarily examine the time evolution of the velocity distribution function, focusing on acceleration driven by the streaming flow. For simplicity, we adopt Model 3, which assumes a constant streaming velocity, to investigate how the streaming flow influences the efficiency of producing suprathermal particles. Figure~\ref{fig:f3} illustrates the effect of streaming velocity on the temporal evolution of the velocity distribution function. In cases with low streaming velocity (i.e., $v_{\text{st}}/v_{\text{th}} = 0$ and $0.1$), the acceleration is primarily isotropic, driven by advection and diffusion. In contrast, for cases with $v_{\text{st}}/v_{\text{th}} \gtrsim 0.5$, the acceleration due to the streaming flow becomes dominant.
Such behavior is qualitatively consistent with in-situ observations of space plasmas, 
where velocity distribution functions often exhibit non-Maxwellian features and asymmetric velocity distributions with enhanced high-energy tails associated with large-scale streaming motions and wave-particle interactions \cite{stverak2009, pierrard2010}. 
Although the present model adopts a simplified one-dimensional and isotropic description, the effective drift term captures the essential role of systematic energization relative to a background flow. 
In this sense, the formation of suprathermal particles in our model can be interpreted as a minimal representation of the processes that may contribute to the development of asymmetric velocity distributions in space plasmas.

To investigate the time evolution of the particle distribution function, particularly with regard to the generation of suprathermal particles, we calculate the fraction of suprathermal particles, defined as:
\begin{equation}
\zeta(t) = \frac{
    \displaystyle \int_{0}^{L} \left[ 
        \int_{-v_{\max}}^{-v_{\text{spt}}} f(x, v, t)\, dv + 
        \int_{v_{\text{spt}}}^{v_{\max}} f(x, v, t)\, dv 
    \right] dx
}{
    \displaystyle \int_{0}^{L} \int_{-v_{\max}}^{v_{\max}} f(x, v, t)\, dv\, dx
},
\label{eq23}
\end{equation}
where $L$ is the size of the spatial domain, and $v_{\text{spt}}$ denotes the threshold velocity above which particles are considered suprathermal. Although $v_{\text{spt}}$ is a free parameter, we adopt $v_{\text{spt}} = 5v_{\text{th}}$ throughout this study.

\begin{figure}[t]
    \centering
    \includegraphics[width=0.7\linewidth]{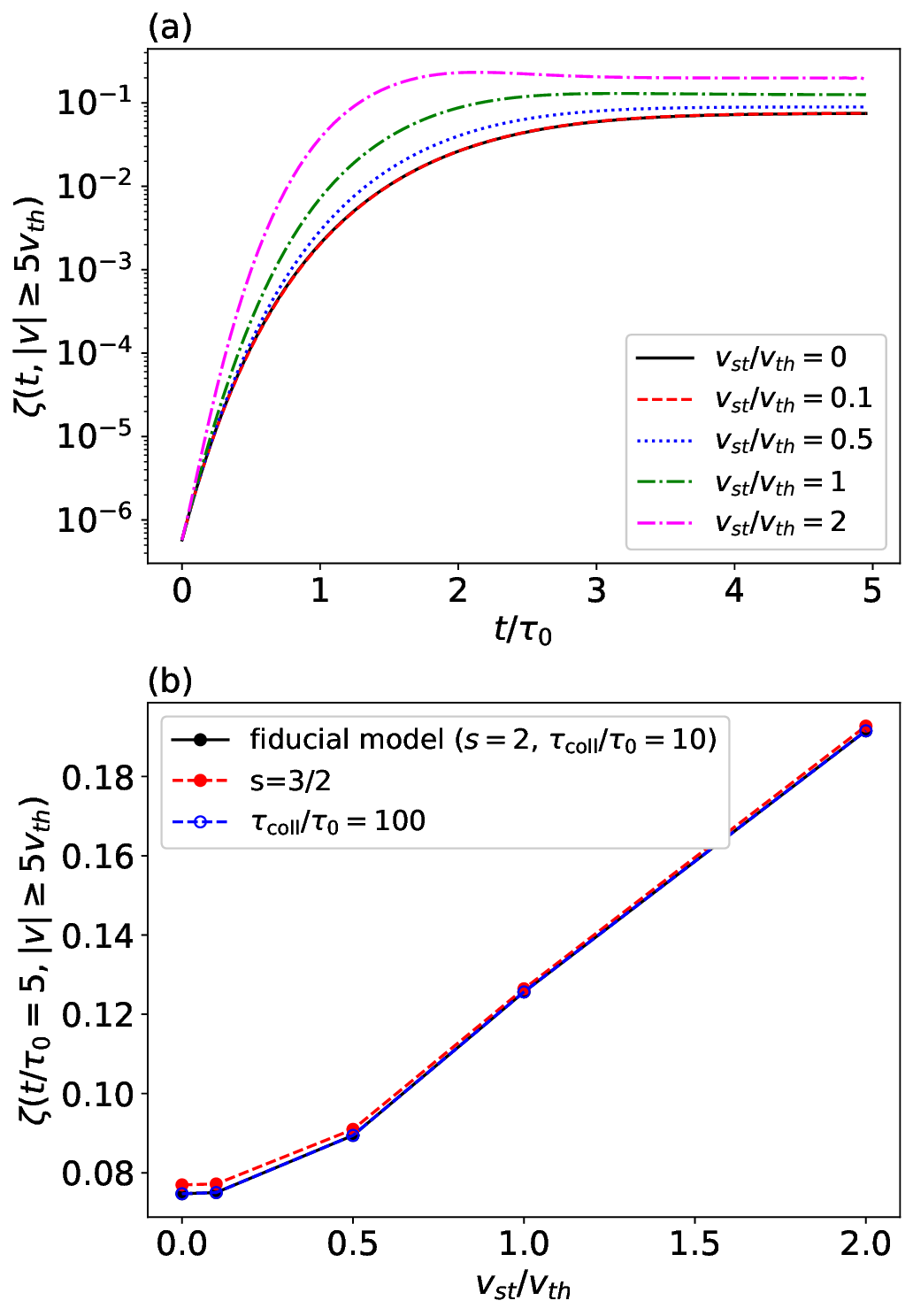}
    \caption{(a) Time evolution of the suprathermal fraction for different values of $v_{\text{st}}$ using the fiducial model ($s=2$, $\tau_{\rm coll} = 10\,\tau_0$). (b) Suprathermal fraction as a function of $v_{\text{st}}$ for the fiducial model, together with models using $s=3/2$ and $\tau_{\rm coll} = 100\,\tau_0$.}
    \label{fig:f4}
\end{figure}

In panel (a) of Figure~\ref{fig:f4}, we present the time evolution of the suprathermal particle fraction for different values of $v_{\text{st}}$. Although an enhancement of the suprathermal fraction is observed on a timescale of order $\tau_0$, the magnitude of $v_{\text{st}}$ governs both the rate of growth and the saturation level of the suprathermal population. In particular, larger values of $v_{\text{st}}$ lead to more rapid energization and higher final suprathermal fractions. 
As shown in panel (b) of Figure~\ref{fig:f4}, the suprathermal fraction increases approximately linearly with increasing $v_{\text{st}}$. We also find that the main trend is largely insensitive to variations in the diffusion scaling and the collisional timescale, indicating the robustness of the result. However, for relatively small streaming velocities ($v_{\text{st}} \lesssim 0.5\,v_{\text{th}}$), diffusion plays a more important role in the early-stage evolution, leading to slight deviations in the suprathermal fraction depending on the model parameters.

\begin{figure}[t]
    \centering
    \includegraphics[width=1\linewidth]{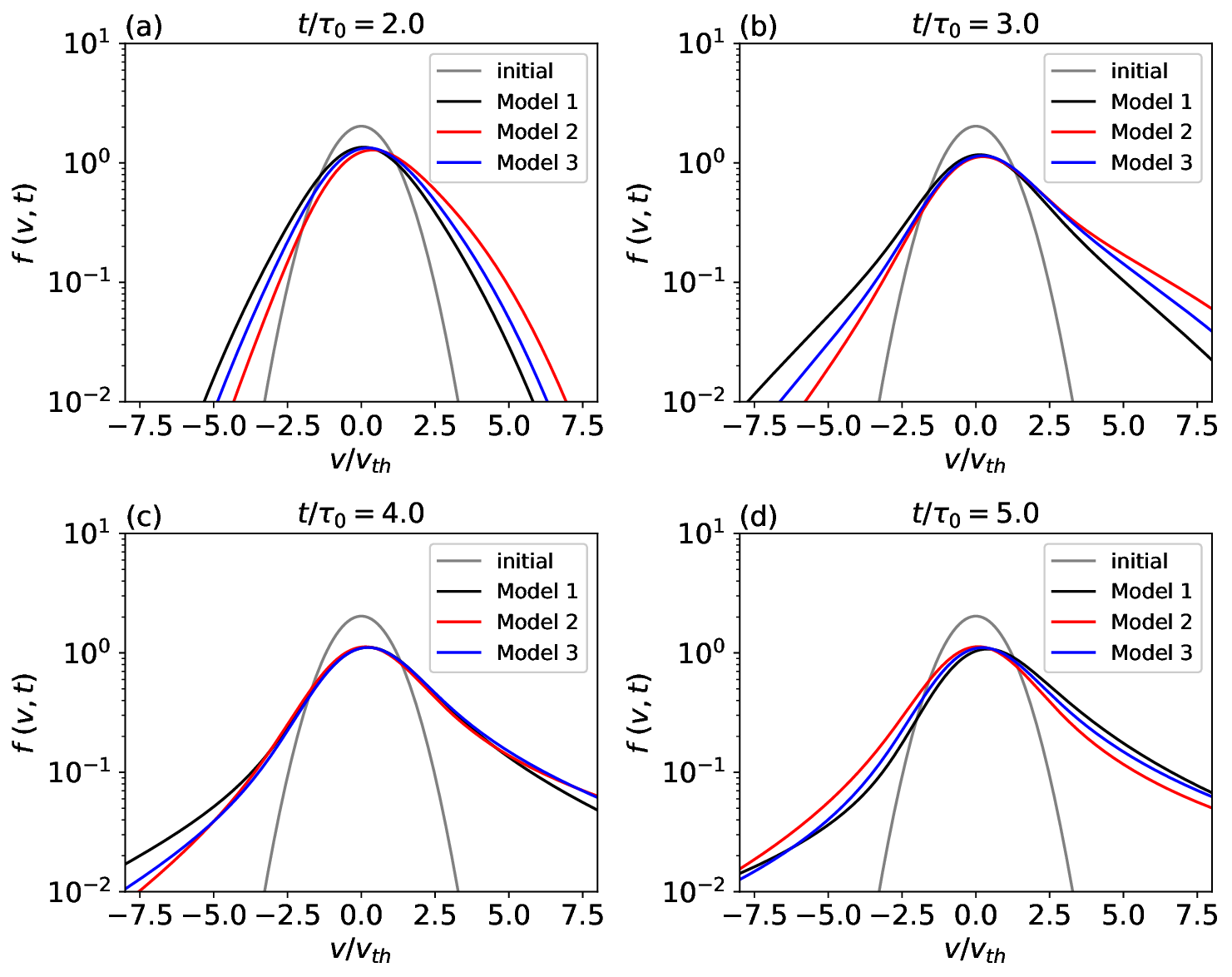}
    \caption{Time evolution of the particle distribution function for three different models. Models 1 and 2 assume accelerating and decelerating streaming flows, respectively, while Model 3 assumes a constant streaming velocity. The fiducial parameters ($s=2$, $\tau_{\rm coll} = 10\,\tau_0$) are adopted for the diffusion and collisional terms. The time-averaged streaming velocity is comparable among the models, with $\langle v_{\rm st} \rangle \approx 0.41\,v_{\rm th}$.}
    \label{fig:f5}
\end{figure}

Next, we examine how the time-dependent form of the streaming velocity influences the evolution of the particle distribution function. The three different models introduced in Section \ref{sec2}, as illustrated in Figure \ref{fig:f5}, are employed for this purpose. During the early evolutionary phase (see panels (a) and (b)), acceleration driven by the streaming flow is most pronounced in Model 2 due to its relatively higher streaming velocity. In contrast, Model 1 exhibits predominantly isotropic acceleration during this phase. The opposite trend emerges later, reflecting the time evolution of the streaming velocity in each model. 
Although the instantaneous acceleration behavior differs between the models, their time-averaged streaming velocities are comparable, with $\langle v_{\text{st}} \rangle \approx 0.41\,v_{\text{th}}$ (see Figure \ref{fig:f2}). As a result, the overall acceleration efficiency shows only a weak dependence on the detailed time variation of the streaming velocity. Rather, it is the time-averaged magnitude of the streaming velocity that primarily governs the fraction of suprathermal particles. This characteristic value of the streaming velocity is also comparable to the typical turbulent velocities expected in the intracluster medium (ICM), where the turbulent Mach number is of order $\mathcal{M} \sim 0.3$--$0.5$ \cite{brunetti2014}. Since the thermal velocity is of the same order as the sound speed in such plasmas, this implies $v_{\text{st}} \sim 0.3$--$0.5\,v_{\text{th}}$, consistent with the representative value adopted in this work. This correspondence suggests that the present framework may provide a useful basis for interpreting suprathermal particle production in galaxy cluster environments.

\begin{figure}[t]
    \centering
    \includegraphics[width=0.7\linewidth]{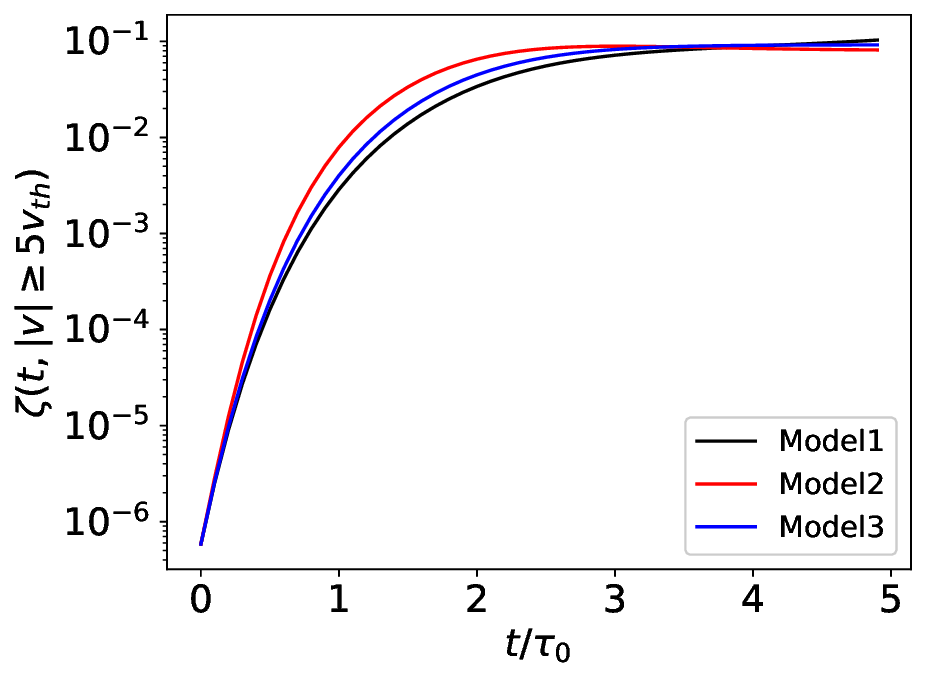}
    \caption{Time evolution of the suprathermal fraction for three different models.}
    \label{fig:f6}
\end{figure}

Figure \ref{fig:f6} shows the time evolution of suprathermal fractions for three different models. We point out that during the early evolutionary phase, the characteristic timescale for producing suprathermal particles critically depends on the time-varying scenario of the streaming velocity. In contrast, the saturated values of suprathermal fractions show only weak sensitivity to the choice of model. These results suggest that while the temporal profile of the streaming velocity governs the initial acceleration process, the system eventually evolves toward a quasi-steady state where the suprathermal particle distribution saturates.

Although space and astrophysical plasmas often undergo multi-phase evolution, including transitions from bursty injection through nonlinear wave-particle interactions to quasi-steady transport, the overall production efficiency of suprathermal particles is largely determined by the time-averaged streaming velocity.
In this context, even under complex and dynamically evolving conditions, observationally or macroscopically constrained mean flow velocities can provide a reliable estimate of the total suprathermal particle yield. 
This reinforces the validity of simplified modeling approaches based on average plasma parameters when interpreting nonthermal particle populations in a wide range of environments.
Consistent with this view, our results support the common simplification adopted in previous studies, where external forces are treated as static fields such as constant electric or gravitational acceleration \cite{pierrard2011, pierrard2013, tang2024, ha2025}. 
Such approximations remain effective for estimating long-term or equilibrium features of suprathermal particle populations. 
However, for investigations targeting transient behavior or short-term bursts in particle acceleration, a time-dependent treatment of streaming flows is essential, as the early-time acceleration rate is governed by the temporal characteristics of the streaming velocity, as demonstrated in Figure \ref{fig:f6}.
This perspective may be particularly relevant in environments such as the solar wind, galaxy cluster plasmas, and relativistic outflows, where large-scale flows and weak collisionality coexist.

\section{Summary}\label{sec4}

In this work, we investigate the time evolution of the particle distribution function by solving the time-dependent Fokker-Planck equation. In addition to turbulent diffusion represented by a phenomenological velocity-space diffusion coefficient, we incorporate the effect of acceleration driven by streaming flow in the background plasma. To capture dynamically evolving environments, a time-dependent streaming velocity is employed to model the background flow.

Our main finding is that the particle acceleration efficiency increases with the magnitude of the streaming velocity, as quantified by the number fraction of suprathermal particles. We further analyze how the specific temporal evolution of the streaming velocity influences the time evolution of the distribution function. 
In cases where the streaming velocity increases from an initially low value, the early phase of particle acceleration is dominated by diffusion. 
As the streaming velocity becomes sufficiently strong, it begins to dominate over the diffusive process, leading to asymmetric acceleration features. 
Conversely, when the streaming velocity decreases from an initially high value, we observe the opposite trend. While the temporal evolution of the streaming velocity affects the intermediate dynamics of particle acceleration, we find that the final acceleration efficiency at saturation depends only weakly on the detailed time profile of the streaming flow. 
Instead, it is primarily determined by the time-averaged streaming velocity. 
These results highlight the importance of considering both the instantaneous and averaged properties of plasma flow when modeling particle acceleration in space and astrophysical environments. 
In particular, the weak dependence of the final acceleration efficiency on the detailed temporal profile of the streaming flow suggests that time-averaged bulk properties provide a robust descriptor of suprathermal particle production. 
This supports the use of simplified effective models in situations where the underlying plasma dynamics are complex and time-dependent.



\section*{Acknowledgements}
We thank to anonymous referees for providing constructive comments to improve the manuscript.


\clearpage

\end{document}